# A Theoretical Mechanism of Szilard Engine Function in Nucleic Acids and the Implications for Quantum Coherence in Biological Systems

F. Matthew Mihelic, MD*

*Lynchburg, Virginia, USA

**Abstract.** Nucleic acids theoretically possess a Szilard engine function that can convert the energy associated with the Shannon entropy of molecules for which they have coded recognition, into the useful work of geometric reconfiguration of the nucleic acid molecule. This function is logically reversible because its mechanism is literally and physically constructed out of the information necessary to reduce the Shannon entropy of such molecules, which means that this information exists on both sides of the theoretical engine, and because information is retained in the geometric degrees of freedom of the nucleic acid molecule, a quantum gate is formed through which multi-state nucleic acid qubits can interact. Entangled biophotons emitted as a consequence of symmetry breaking nucleic acid Szilard engine (NASE) function can be used to coordinate relative positioning of different nucleic acid locations, both within and between cells, thus providing the potential for quantum coherence of an entire biological system. Theoretical implications of understanding biological systems as such "quantum adaptive systems" include the potential for multi-agent based quantum computing, and a better understanding of systemic pathologies such as cancer, as being related to a loss of systemic quantum coherence.

## INTRODUCTION

It has been theorized that quantum decision functions are performed by biological systems, but the mechanisms of such functions have not been heretofore characterized. Three requirements of quantum logic functions are superposition of information, a quantum gate, and quantum entanglement, and these three elements theoretically exist in the nucleic acids of biological systems to effect logical decision functions that decrease entropy and provide for system coherence.

## ENTROPY

Claude Shannon's theory of information views data to be high in entropy, while the conduit that carries the data is considered to be low in entropy [1]. A molecule that is traveling through the conduit of intracellular or extracellular fluid could thus be considered to be data that is high in entropy. That entropic data molecule might be "recognized" by a congruently matched cellular recognition instrument (e.g. binding protein), and with that recognition and consequent binding, the entropy of the data molecule could then be transferred to the recognition instrument, and that recognition instrument (or binding protein) could then be considered to be an entropic data molecule in its own right. That recognition instrument (which is now an entropic data molecule in its own right) could then encounter and bind with an appropriate portion of nucleic acid source code and transfer that entropy to the nucleic acid.

If that data molecule/recognition instrument were to encounter a congruently matched section of a nucleic acid molecule, it could then bind to that section in a manner analogous to atoms or molecules fitting into a crystalline lattice. The entropy of the data molecule would then become shared within the entire complex consisting of the nucleic acid and the data molecule. In Leo Szilard's interpretation of the "Maxwell's demon" thought experiment, entropy can be transformed into useful work by a homunculus inside the mechanism of a theoretical machine that uses information about a particle in motion to convert the particle's "microscopic" entropic particular motion into the mechanical movement of a piston [2]. The essence of such a "Szilard engine" is that information can be used to convert the energy associated with entropy into useful work at a "microscopic" level, and according to Rolf Landauer all information is physical [3], and so a nucleic acid might be seen as being such a "Szilard engine" that is literally and physically built out of information. In effect this puts the homunculus of the "Szilard engine" both inside and outside the machine at the same time, and this puts the information on both sides of the theoretical machine as well. It would also follow from this that the function of such a "nucleic acid Szilard engine" would be logically reversible if the information from the data molecule could be retained in the mechanical degrees of freedom of the geometric configuration of the nucleic acid molecule.

Leo Szilard had theorized a mechanism by which information can be used to harness entropy to do useful work at a "microscopic" level, and a nucleic acid could theoretically operate as such a "Szilard engine" to transform the entropy of the molecular data that it recognizes (i.e. is coded for) into the useful work of geometric reconfiguration (within the mechanical degrees of freedom) of the nucleic acid molecule, because it is physically constructed out of the information necessary to reduce the entropy of the data that it encounters. Environmental interactions could thus bring about a geometric reconfiguration of a nucleic acid if that nucleic acid recognizes what it is interacting with, through encoding of such recognition in the nucleic acid sequence. Such a "nucleic acid Szilard engine" (NASE) thus functions to reconfigure a nucleic acid in response to environmental interaction, as per coded recognition of the environmental stimulus (e.g. entropic data molecule).

## QUANTUM ENTANGLEMENT

Photons are known to be both emitted and absorbed by nucleic acids, and there is a mandatory emission of photons during a quantum search [4]. Because the function of the NASE essentially effects a quantum measurement or search, there would necessarily be such a mandatory photon emission in such a symmetry breaking event. Entangled photons thus generated could coordinate relative geometric reconfiguration between nucleic acid segment locations through coherent signaling from between segments, as a response to "recognition" of a piece of molecular data by a particular congruent nucleic acid segment. Quantum information could thus be retained in the geometric configuration of the nucleic acid source code. Signaling and relative geometric reconfiguration between nucleic acid positions could occur within (or between) a cell's nucleic acid(s), or between the nucleic acids of two different cells within the same system. This quantum information is the dynamic information through which the NASE mechanism ensures that the relative nucleic acid geometric reconfiguration only takes place via energy transfer between states entangled with the system.

## QUANTUM GATE

The "nucleic acid Szilard engine" (NASE) function of a nucleic acid allows for geometric reconfiguration of a nucleic acid molecule in response to the transfer of information entropy from a data molecule (potentially through a recognition instrument, e.g. binding protein) to the nucleic acid. The entropy from the data molecule is converted into useful work and is thus lowered (with respect to the source code of the system) in a manner analogous to a catalyst lowering entropy in a chemical reaction. Parrando said that any Szilard engine function is a symmetry break, so NASE function could be considered to be the "symmetry breaking" result of a "measurement" constituting a quantum decision in a quantum search [5].

The mechanism of the NASE is logically reversible because the information from the data molecule is retained in the mechanical degrees of freedom of nucleic acid geometric configuration, and as such can constitute a quantum gate through which two qubits can connect, and which can then be used to perform quantum logic. A qubit can be considered to be anything that is "multi-state" or has superimposed information (e.g. a multi-dimensional source code), and so two agents (e.g. cells) containing multi-state geometrically reconfigurable source codes (e.g. DNA) can interact at the level of their source codes via a quantum gate (e.g. NASE) to lower entropy and thus perform quantum logic. The nucleic acid molecule can thus be seen as providing superimposed information as qubits consisting of the geometric configuration of the molecule, and also as providing the quantum gate consisting of the NASE mechanism.

## QUANTUM COHERENCE

The coherent energy transfer for relative geometric reconfiguration of nucleic acid source code takes place through a theoretical quantum gate provided by the NASE mechanism, and it is through such a quantum gate that quantum logic, and consequently a quantum search, can be performed. There is significant indication that biophotons can affect cells other than the one from which they originate [6] [7], and so this would imply the potential for quantum coherence of a biological system that is based upon each cell having a different geometric configuration of an identical linear DNA source code, and in which the entirety of such a system would consequently be available for quantum search via entangled biophotons.

## THEORETICAL IMPLICATIONS

Schrödinger had theorized that quantum mechanical processes are an essential component of a definition of "life", and that the adaptive processes necessary for "life" involve "quantum leaps" of electrons between energy states within the covalent chemical bonds of the "aperiodic crystal" that carries the genetic information of a biological system [8]. The coherent stability of such an "aperiodic crystal" was found in DNA, and the NASE provides a mechanism by which a nucleic acid can act as an information-carrying catalyst to lower the entropy of a system in such a manner that in effect allows data to "crystallize" around it. This provides insight into how the conscious observer can cause the "collapse of the wave function" through the nucleic acid Szilard engine's symmetry breaking measurement or quantum decision. A measurement is a comparison to a source code (e.g. DNA or RNA), and it is such a measurement that can provide a molecular representation of the objective physical world to the subjective mind as the nucleic acid location, and the moment, of the demarcation (or "schnitt") between what von Neumann referred to as "psycho-physical parallelism" [9].

With this understanding of the nucleic acid basis of the quantum coherence of biological systems, what have previously been thought of as biological "complex adaptive systems" can now be better understood as "_quantum_ adaptive systems". Such theoretical conceptualization can be envisioned to lead to new "disruptive" technologies. For instance, multi-agent quantum computing based upon molecular geometric conformational changes can be thought of as a more viable design for quantum computing, than one that is based upon fleeting artificial stabilization of the coherence of subatomic particles. Also, a better understanding of cancer as a loss of systemic quantum coherence should lead to new efficacious therapies.

## CONCLUSION

Nucleic acids theoretically possess a Szilard engine function that can convert the energy associated with the Shannon entropy of molecules for which they have coded recognition, into the useful work of geometric reconfiguration. Such a mechanism is logically reversible because it is physically constructed out of the information necessary to reduce such entropy, and because information is retained in the geometric degrees of freedom of the nucleic acid molecule a quantum gate is formed through which multi-state nucleic acid qubits can interact. Entangled biophotons emitted as a consequence of symmetry breaking Szilard engine function can be used to coordinate relative positioning of different nucleic acid locations, both within and between cells, thus providing the potential for quantum coherence of a biological system.

The nucleic acid Szilard engine (NASE) model of quantum coherence in biological systems is consistent with Schrödinger's predictions of biological quantum coherence carried through the genetic material of the cell nucleus, and it provides explanation of the observer effect in the double slit experiment that is consistent with von Neumann's "psycho-physical parallelism". This model reveals the basic mechanism of the quantum complexity of biological systems, and can form the basis of new directions in quantum computing and evaluation of complex systemic pathologies.